\documentclass[conference]{IEEEtran}
\usepackage[normalem]{ulem}
\usepackage{xspace}
\usepackage[hidelinks]{hyperref}
\usepackage{tikz}
\usepackage{fancyvrb}

\usepackage{tgcursor}
\usepackage{pifont}
\usepackage{cleveref}
\usepackage{subfig}
\usepackage{multirow}
\usepackage{paralist}
\usepackage{wasysym}
\usepackage{mdframed}
\usepackage{tcolorbox}

\usepackage[frozencache=true,cachedir=minted-cache]{minted}

\definecolor{beaublue}{HTML}{e2e6ed}
\colorlet{shadecolor}{beaublue}
\tcbset{
	colback=beaublue,
	boxrule=0.5pt,
	boxsep=1pt,
	left=3pt,
	right=3pt,
	top=3pt,
	bottom=3pt,
	sharp corners,
	colframe=white,
}

\newcommand{\insight}[1]{
	\begin{tcolorbox}
		\textbf{Insight:} #1
	\end{tcolorbox}
}

\setminted{fontsize=\footnotesize}

\newcommand\customtt[1]{{\small \texttt{#1}}}
\newcommand\tinytt[1]{{\scriptsize \texttt{#1}}}

\usepackage{color}
\usepackage{tcolorbox}

\definecolor{lightgrey}{RGB}{244,244,244}
\definecolor{darkgrey}{RGB}{99,99,99}

\newcommand{\sys}{ConfFuzz\xspace}

\newcommand{\urlsys}{\url{https://conffuzz.github.io}\xspace}
\newcommand{\nginxcivurl}{\url{https://github.com/conffuzz/conffuzz-ndss-data/blob/main/docs/nginx.md}\xspace}
\newcommand{\smallurlsys}{\footnotesize \urlsys}

\newcommand{\numberofapps}{25\xspace}
\newcommand{\numberoflibs}{36\xspace}
\newcommand{\numberofframeworks}{12\xspace}
\newcommand{\numberofframeworksthatconsidercivs}{2\xspace}
\newcommand{\numberofscenarios}{39\xspace}
\newcommand{\numberofscenariosfromliterature}{16\xspace}
\newcommand{\numberofcrashespriordeduplication}{15,371\xspace}
\newcommand{\numberofcrashes}{629\xspace}
\newcommand{\numberofsafeboxcrashes}{250\xspace}
\newcommand{\numberofsandboxcrashes}{379\xspace}
\newcommand{\numberofbugswithmultipleimpact}{224\xspace}
\newcommand{\sumofimpacts}{999\xspace}

\newcommand{\numberofscenarioswithoutcrashes}{3\xspace}

\newcommand{\percentageofarbitraryreadvulnerabilities}{70}
\newcommand{\percentageofarbitrarywritevulnerabilities}{66}

\newcommand{\CIV}{CIV\xspace}
\newcommand{\CIVfull}{Compartment-Interface Vulnerability\xspace}
\newcommand{\CIVs}{CIVs\xspace}
\newcommand{\CIVsfull}{Compartment-Interface Vulnerabilities\xspace}

\usepackage{xcolor}

\newcommand*\BC[1]{\tikz[baseline=(char.base)]{
	\node[shape=circle,draw,inner sep=0.15pt] (char) {\textcolor{black}{#1}};}}

\begin{document}

\title{Assessing the Impact of Interface Vulnerabilities in Compartmentalized Software}

\newcommand{\manchester}{$\dagger$}
\newcommand{\rice}{$\ddagger$}
\newcommand{\unikraft}{\tiny$\infty$}
\newcommand{\upb}{\tiny$\nabla$}

\author{\IEEEauthorblockN{Hugo Lefeuvre\textsuperscript{\manchester}, Vlad-Andrei Bădoiu\textsuperscript{\upb}, Yi Chien\textsuperscript{\rice}, Felipe Huici\textsuperscript{\unikraft}, Nathan Dautenhahn\textsuperscript{\rice}, Pierre Olivier\textsuperscript{\manchester}}
\IEEEauthorblockA{\emph{\textsuperscript{\manchester}The University of Manchester, \textsuperscript{\upb}University Politehnica of Bucharest, \textsuperscript{\rice}Rice University, \textsuperscript{\unikraft}Unikraft.io}}}

\date{}

\IEEEoverridecommandlockouts
\makeatletter\def\@IEEEpubidpullup{5.5\baselineskip}\makeatother
\IEEEpubid{\parbox{\columnwidth}{
    Network and Distributed System Security (NDSS) Symposium 2023\\
    27 February - 3 March 2023, San Diego, CA, USA\\
    ISBN 1-891562-83-5\\
    https://dx.doi.org/10.14722/ndss.2023.24117\\
    www.ndss-symposium.org
}
\hspace{\columnsep}\makebox[\columnwidth]{}}

\maketitle

\begin{abstract}

Least-privilege separation decomposes applications into compartments limited to
accessing only what they need. When compartmentalizing existing software, many
approaches neglect securing the new inter-compartment interfaces, although what
used to be a function call from/to a trusted component is now potentially a
targeted attack from a malicious compartment.  This results in an entire class
of security bugs: Compartment Interface Vulnerabilities (CIVs).

This paper provides an in-depth study of CIVs. We taxonomize these issues and
show that they affect all known compartmentalization approaches. We propose
\sys, an in-memory fuzzer specialized to detect CIVs at possible compartment
boundaries.  We apply \sys to a set of \numberofapps popular applications and
\numberoflibs possible compartment APIs, to uncover a wide data-set of
\numberofcrashes vulnerabilities.  We systematically study these issues, and
extract numerous insights on the prevalence of CIVs, their causes, impact, and
the complexity to address them.  We stress the critical importance of CIVs in
compartmentalization approaches, demonstrating an attack to extract isolated
keys in OpenSSL and uncovering a decade-old vulnerability in sudo. We
show, among others, that not all interfaces are affected in the same way, that API size is
uncorrelated with CIV prevalence, and that addressing interface
vulnerabilities goes beyond writing simple checks. We conclude the paper with
guidelines for CIV-aware compartment interface design, and appeal for more
research towards systematic CIV detection and mitigation.

\end{abstract}

\section{Introduction}
\label{sec:introduction}

The principle of least privilege has guided the design of safe computer systems
for over half a century by ensuring that each unit of trust in a system can
access only what it truly needs to fulfill its duties: in this way, system
designers can proactively defend against unknown
vulnerabilities~\cite{Watson2015}. Software compartmentalization is a prime
example where unsafe, untrusted, or high-risk components are
isolated to reduce the damage they would cause should they be
compromised~\cite{Saltzer1975}.

Recent years have seen the appearance of an increasingly large number of new
isolation mechanisms~\cite{Costan2016, ARMTrustZone2009, ARMMorello2020,
Watson2015, Schrammel2020, Park2019} that enable fine-grained compartmentalization.
This resulted in compartmentalization works targeting finer
and finer granularities, such as libraries~\cite{Wu2012,
VahldiekOberwagner2019, Hedayati2019, Narayan2020, Schrammel2020, Liu2017,
Bauer2021, Sartakov2021, Lefeuvre2022, Almatary2022}, modules~\cite{Im2021, Almatary2022,
Schrammel2022}, files~\cite{Almatary2022}, and even functions/blocks of
code~\cite{Gudka2015, Wanninger2022, Sung2020, Agadakos2022}.  In that context,
major attention was dedicated to compartmentalizing existing code, since
rewriting software from scratch to work in a compartmentalized manner is costly
and complex~\cite{Gudka2015}.  With recent developments on compiler-based
compartmentalization, frameworks offer to apply isolation at arbitrary
interfaces for a low to non-existent porting cost~\cite{Wu2012, Bauer2021,
Liu2017, Agadakos2022}.

Unfortunately, breaking down applications into compartments means that control
and data dependencies through shared interfaces create new classes of
vulnerabilities~\cite{VanBulck2019}: in order to provide safe
compartmentalization, it is not only necessary to ensure spatial memory
isolation but also to design interfaces with distrust in mind. For example,
objects passed through APIs can be corrupted to launch confused deputy
attacks~\cite{Mao2011, Hu2015}, data structures can be manipulated to control
execution or leak data through Iago attacks~\cite{Checkoway2013,
Cui2022}, called components can modify return values or indirectly access
shared data structures to launch new forms of exploit, etc.

Even though interface-related vulnerabilities (denoted \emph{\CIVsfull} / \CIVs
in this paper) were previously identified to various extents in the
literature~\cite{Mao2011, Checkoway2013, Hu2015, VanBulck2019}, almost
all modern compartmentalization frameworks~\cite{Wu2012,
VahldiekOberwagner2019, Hedayati2019, Schrammel2020, Liu2017, Koning2017,
Park2019, Bauer2021, Sartakov2021, Sung2020, Lefeuvre2021, Lefeuvre2022,
Agadakos2022} neglect the problem of securing interfaces, and rather focus on
transparent and lightweight spatial separation.  Since \CIVs are already
problematic for interfaces hardened from the ground up (e.g., the system call
API~\cite{LINUX_CONFUSED_DEPUTY_1, Checkoway2013}) with well-defined
trust-models (kernel/user), their impact on safety is likely to be even greater
when considering arbitrary interfaces and trust models that materialize when
compartmentalizing existing software that was not designed with the assumption
of hostile internal threats. Worse still, the complexity of safeguarding
interfaces increases as more fine-grain components are targeted.

Beyond this lack of consideration, \CIVs remain misunderstood; we ask the
following research questions: \emph{how widespread are \CIVs when
compartmentalizing unmodified applications? What are the API design patterns
leading to them? What is the concrete impact of \CIVs on the safety guarantees
brought by compartmentalization, and what is the complexity of addressing
them?} In order to achieve \CIV mitigations that are generic and principled, we
stress the need to formalize and quantify the problem.

This paper provides an in-depth study of \CIVs.  We taxonomize \CIVs into a
coherent framework, and systematize existing efforts to address them,
highlighting categories that need attention in future research. In order to
study existing \CIVs in real-world scenarios, we propose \sys, an in-memory
fuzzer specialized to detect \CIVs at possible compartment boundaries.  \sys
automatically explores the complexity of compartment interfaces by exposing
data dependencies leading to vulnerabilities. Contrary to existing fuzzers,
that inject malformed data in a single direction (e.g., a library), \sys can
show the degree to which data flowing through an interface can be manipulated
to harm either direction of a cross-compartment call.  We apply \sys to a
corpus of \numberofscenarios compartmentalization scenarios, many of which
previously proposed as use-cases of \numberofframeworks existing research and
industry frameworks. We uncover a wide data-set of \numberofcrashes potential
vulnerabilities\footnote{We open-sourced code and data-set: \smallurlsys}. We
systematically study these issues, extracting numerous insights on the
prevalence of \CIVs, their causes, impact, and the complexity to address them.

At the highest level, our results confirm how important the problem of \CIVs
should be to compartmentalization research: in many cases, \CIVs seriously
reduce or even fully negate all benefits of compartmentalization, and that even when
the interface is extremely simple: we demonstrate an attack to extract isolated
keys in OpenSSL, a common application of compartmentalization, and a
decade-old vulnerability in sudo's authentication API.  Beyond this, we note
the following high-level insights: 1) \CIVs are present in almost all existing
interfaces, but at significantly varying degrees: for instance module
APIs are much more vulnerable than library interfaces,
and some interfaces are entirely CIV-free; 2)
the complexity of objects crossing the interfaces imports more than the size of
the API itself, and most of an API's \CIVs can often be tracked down to a
handful of objects; 3) fixing \CIVs goes further than writing a few checks, and
often requires reworking interfaces and partially redesigning
existing software. We conclude with an appeal for more research towards
systematic \CIV detection and mitigation, hoping that this study can encourage
future works to consider the issue of interfaces.

To sum up, this paper makes the following contributions:
\begin{compactitem}
    \item A systematization and taxonomy of \CIVs and existing efforts to address
        them (\S\ref{sec:taxonomy}).

    \item \sys, an in-memory fuzzer that automatically detects \CIVs in existing
	software at arbitrary interfaces (\S\ref{sec:design-and-implementation}).

    \item A systematic study of the \CIVs found by \sys applied to \numberofscenarios
        real-world application compartmentalization scenarios, backing insights with
	concrete data (\S\ref{sec:evaluation}).

    \item A series of interface design guidelines intended to ease the
        development/adaptation of new/existing interfaces with
	compartmentalization in mind (\S\ref{sec:discussion}).
\end{compactitem}

\section{Motivation}
\label{sec:motivation}

The problem of secure interface design is not new~\cite{Hardy1988}. The Linux
system call interface, for example, is the result of years of organic evolution towards a
strong boundary that preserves the integrity of the kernel in the presence of
untrusted applications. Alas, designing strong interfaces in an adversarial
context is notably hard: interface-related vulnerabilities are still regularly
reported against the system call interface~\cite{LINUX_CONFUSED_DEPUTY_1, Koschel2020, Kemerlis2014}, even
after decades of hardening.  The task is even harder when assuming
mutual distrust; the system call API, to take the same example,
is notably weak at protecting the application from the
kernel~\cite{Checkoway2013, Cui2022}, and requires extensive shim
interfaces~\cite{OE_SDK, Tsai2017, Priebe2019, ASYLO, FORTANIX_EDP} to sanitize
untrusted inputs and outputs.

Modern compartmentalization frameworks enable users to easily enforce spatial
and temporal separation between components of existing software.  Typically,
code is either \emph{sandboxed}, where a software component prone to subversion
is restricted from accessing the rest of the system (e.g., image processing
libraries), \emph{safeboxed}, where sensitive data is only accessible to a
component while maintaining high privilege (e.g., libssl), or \emph{separated}
into mutually distrusting subsystems~\cite{Lampson1974}. Depending on domain
crossing frequencies, compartmentalization promises good vulnerability
containment at a modest cost~\cite{Watson2015}.

Unfortunately, as shown by previous works~\cite{Hu2015, VanBulck2019,
Narayan2020} and highlighted in this paper, simply isolating software
components is not enough: if cross-compartment interfaces have not been
designed as trust boundaries (e.g., when compartmentalizing existing software),
a wide range of \emph{\CIVsfull} (\CIVs) arise. Reasoning about the safety of
an interface is complex due to data dependencies exposed through the use of
that interface by actors that previously belonged to a single trust domain, but
under compartmentalization distrust each other. That complexity increases with
that of interface-crossing data flows. \CIVs arise when developers would like
to avoid trust in a component, and encompass traditional confused
deputies~\cite{Hardy1988}, Iago
vulnerabilities~\cite{Checkoway2013}, or Dereferences Under Influence
(DUIs)~\cite{Hu2015}. We define CIVs more formally in \S\ref{sec:taxonomy}.

\begin{listing}
  \begin{minted}{C}
// ImageMagick callback exposed to libghostscript
static int MagickDLLCall GhostscriptDelegateMessage(
  void *handle, const char *message, int length) {

  /* CIV: unchecked dereference/usage of
   * sandbox-provided pointer/bounds information */
  (void) memcpy(*handle, message, (size_t) length);
  (*handle)[length] ='\0';
} /* ... abbreviated / simplified ... */
  \end{minted}
  \caption{
    ImageMagick callback lets libghostscript perform arbitrary writes outside the sandbox.
  }
  \label{fig:im-arbitrary-write}
\end{listing}

Take the example of library sandboxing in ImageMagick as done by the
Compiler-Assisted Library Isolation (Cali)~\cite{Bauer2021} framework. Here,
libghostscript is sandboxed because it is notoriously prone to
high-impact vulnerabilities. Cali automatically sandboxes the library by
applying compiler-based techniques to detect data shared between the
application and the library, and place them in a shared memory region, before running
application and library in separate processes.  When the
application needs to execute a function of the library, Cali performs the
function call in the library compartment.  Whenever the library
needs to execute an application callback, Cali executes the callback in the
application compartment.

This approach might seem sufficient: it makes it harder for attackers to escape
the sandboxing of libghostscript. In practice, however, as shown in
\Cref{fig:im-arbitrary-write}, ImageMagick exposes a callback to libghostscript
that allows the untrusted library to perform arbitrary writes in the
application's compartment as often as it wants and at any time, negating
spatial isolation entirely. This vulnerability, identified by our tool \sys, is
caused by ImageMagick (victim compartment) dereferencing sandbox-provided
pointers (\customtt{handle} and \customtt{message}) and bounds information
(\customtt{length}) without sufficient checking.

Even though \CIVs particularly affect new fine-grain compartmentalization
frameworks such as Cali, they are not a specificity of these frameworks; as we
show in \S\ref{sec:evaluation}, \CIVs also affect long-standing,
production-grade sandboxing approaches such as the worker/master separation in
Nginx.

Clearly, while a strong compartmentalization framework capable of reliably
enforcing spatial and temporal isolation is necessary, it is insufficient to
offer tangible security benefits: software must also be adapted to fit distrust
scenarios by vetting information that crosses compartment interfaces.

In the remainder of this paper, we propose the first systematization and
taxonomy of \CIVs and existing defenses, introduce \sys, a tool to
automatically detect \CIVs at potential compartmentalization boundaries, and
use it to provide an in-depth study of real-world \CIVs found with \sys.

\section{\CIVsfull}
\label{sec:taxonomy}

In this section, we provide the first definition and taxonomy of \CIVs, along
with a systematic review of existing defenses and their shortcomings. We define
three main classes of \CIVs, subdivided in a total of 8 subclasses. We relate
each subclass with existing mitigations and discuss their limits, summarized in
\Cref{tab:classification}.  Here we use the term \emph{corrupted} to refer to data
voluntarily malformed by a malicious compartment.

\begin{tcolorbox}
A \emph{\CIVfull} (\CIV) is an instance of the general confused
deputy~\cite{Hardy1988} problem, where a compartment is its own deputy, and
fails to adequately vet the use of the interface it exposes to other
compartments, as well as its usage of other compartments' interfaces.
\end{tcolorbox}

Malicious compartments can leverage a \CIV to mount data and control-based
attacks, confusing a victim compartment into leaking and altering its private
data and addresses, executing code, etc.  Many \CIVs arise due to incorrect or
missing checks and sanitization of data flowing through the interface; our
taxonomy provides a comprehensive list of causes. Type
confusion~\cite{Haller2016}, DUIs~\cite{Hu2015}, and Iago~\cite{Checkoway2013}
are all part of the \CIV spectrum. For instance, Iago vulnerabilities are CIVs
at the system-call boundary, and DUIs match DC1 and DC2
(\S\ref{par:DUIs}).

\begin{table*}[]
\centering
\caption{Compartmentalization mechanisms and frameworks that consider certain CIV
	classes. A \CIRCLE{} indicates that a CIV class is fully addressed; a \LEFTcircle{}
	indicates a \emph{partially} addressed CIV class; a \Circle{}
	means fully vulnerable. An asterisk \textbf{*} indicates that the fix is not
	generic: the method makes assumptions about the source code being
	compartmentalized, or the use-case.
}

{
\setlength{\tabcolsep}{5pt} 


\scriptsize
\begin{tabular}{|l|c|c|c|c|c|c|c|c|c|}
\hline
\emph{Mitigation Approach \textbackslash{} CIV Class} & \emph{DL1/Exp.Addr} & \emph{DL2/Exp.Dat} & \emph{DC1/Corr.Pt} & \emph{DC2/Corr.Ind} & \emph{DC3/Corr.Obj} & \emph{TV1/API.Ord} & \emph{TV2/Corr.Sync} & \emph{TV3/Race} \\ \hline\hline
\textsc{Type-Based Checks: RLBox}~\cite{Narayan2020} & \phantom{\textbf{*}}\LEFTcircle\textbf{*} & \Circle & \phantom{\textbf{*}}\CIRCLE\textbf{*}& \LEFTcircle & \phantom{\textbf{*}}\LEFTcircle\textbf{*} & \phantom{\textbf{*}}\LEFTcircle\textbf{*} & \Circle  & \Circle \\ \hline
\textsc{Hardware Capabilities: CHERI}~\cite{Watson2015} & \Circle & \Circle & \phantom{\textbf{*}}\CIRCLE\textbf{*} & \phantom{\textbf{*}}\CIRCLE\textbf{*} & \phantom{\textbf{*}}\LEFTcircle\textbf{*} & \Circle & \Circle & \Circle \\ \hline
\textsc{Undefined Mem. Sanitizers}~\cite{Stepanov2015} & \LEFTcircle & \LEFTcircle & \Circle & \Circle & \Circle & \Circle & \Circle & \Circle \\ \hline
\textsc{Bounds-Checking Techniques}~\cite{Szekeres2013} & \Circle & \Circle & \Circle & \phantom{\textbf{*}}\CIRCLE\textbf{*} & \phantom{\textbf{*}}\LEFTcircle\textbf{*} & \Circle & \Circle & \Circle \\ \hline
\textsc{ASLR-Guard~\cite{Lu2015}, FG-ASLR~\cite{Kil2006}} & \LEFTcircle & \Circle & \Circle & \Circle & \Circle & \Circle & \Circle  & \Circle \\ \hline
\textsc{Annot. + DF-Analysis: SOAAP}~\cite{Gudka2015} & \Circle & \LEFTcircle & \Circle & \Circle & \Circle & \Circle & \Circle  & \Circle \\ \hline
\textsc{Pointer Authentication}~\cite{Mashtizadeh2015, Liljestrand2019} & \Circle & \Circle & \phantom{\textbf{*}}\CIRCLE\textbf{*} & \Circle & \Circle & \Circle & \Circle & \Circle \\ \hline
\textsc{API Sem. Sanitization: APISan}~\cite{Yun2016} & \Circle & \Circle & \Circle & \Circle & \Circle & \LEFTcircle & \Circle & \Circle \\ \hline
\textsc{TOCTTOU Protection: Midas}~\cite{Bhattacharyya2022} & \Circle & \Circle & \Circle & \Circle & \Circle & \Circle & \Circle & \phantom{\textbf{*}}\CIRCLE\textbf{*} \\ \hline
\end{tabular}
\label{tab:classification}
}
\end{table*}

\subsection{Cross-Compartment Data Leakage (DL)}
\label{subsec:info-exposure}

\paragraph{DL1: Exposure of Addresses}

A victim compartment may leak compartment-internal memory addresses, allowing
an attacker, among others, to break ASLR in the victim and locate critical objects.
In \Cref{fig:im-arbitrary-write}, address leaks may help
libghostscript to know where to point \customtt{handle} to.  DL1 can stem from
interface-crossing uninitialized data structures/fields and compiler-added
padding~\cite{Osterlund2019}, as well as data over-sharing between
compartments. RLBox~\cite{Narayan2020} proposes to address DL1 with pointer
swizzling~\cite{Wilson1991}, ensuring that interface-crossing pointers can only
address the sandbox. Generalized to arbitrary compartmentalization
scenarios (e.g., by forcing interface-crossing pointers to address shared
regions), this solves leaks due to oversharing, but does not address leaks due
to uninitialized memory or padding.  A near-complete protection can be achieved
by combining RLBox with uninitialized memory use detectors such as
MSan~\cite{Stepanov2015}.  RLBox is not generic, as it requires strong types
which are not available in all languages (e.g., C), and non-trivial manual
refactoring that forbids certain C/C++ idioms. More generic but weaker
protection can be achieved with ASLR hardening techniques~\cite{Lu2015,
Kil2006}.

\paragraph{DL2: Exposure of Compartment-Confidential Data}

A victim compartment may leak compartment-confidential data to a malicious
compartment. The impact depends on the nature of the leakage; typical targets
include cryptographic secrets, or user data. Leaks stem from over-sharing, as
well as uninitialized shared objects containing data from previous allocations.
SOAAP~\cite{Gudka2015} proposes to address DL2 with manually annotated sensitive
data, leveraging data-flow analysis to guarantee that annotated objects never
cross trust boundaries. This approach does not prevent leakages due to
uninitialized memory/padding, and is notoriously prone to human error: past
attempts at compartmentalizing OpenSSL failed because of misidentification of
private data~\cite{Reisinger2014}. Similarly to DL1, more complete protection
can be achieved by combining SOAAP with uninitialized memory use detectors.

\subsection{Cross-Compartment Data Corruption (DC)}
\label{par:DUIs}

\paragraph{DC1: Dereference of Corrupted Pointer}

A victim compartment may dereference a pointer corrupted by a malicious
compartment. The impact spans that of all classical spatial vulnerabilities:
malicious actors may gain read, write, or execute capabilities in the victim's
context, or cause Denial of Service (DoS). In
\Cref{fig:im-arbitrary-write}, corrupted pointers \customtt{handle} and
\customtt{message} grant the untrusted compartment full write
permissions.  DC1 vectors include shared objects, cross-compartment function
call arguments and return values.  RLBox~\cite{Narayan2020} proposes to
address DC1 with pointer swizzling~\cite{Wilson1991}, with the same limitations
as in DL1.  Hardware memory capabilities such as CHERI~\cite{Watson2015}
address DC1 by making it impossible to forge pointers. Although promising, this
technology is still at a research prototype stage~\cite{ARMMorello2020}.  Its
protection is not generic, requiring porting/annotations to fully address DC1, leaving certain C/C++ idioms
unsupported. Generally, pointer authentication techniques
like ARM PA~\cite{Liljestrand2019} can also address DC1, but may require
porting in a compartmentalized context.

\paragraph{DC2: Usage of Corrupted Indexing Information}

A victim compartment may use indexing information (size, offset, index)
corrupted by a malicious compartment. The impact includes DoS, and that of
buffer overflows, and underflows, depending on the context. Typical vectors
are, similarly to DC1, shared data, cross-compartment function call arguments,
and return values. RLBox~\cite{Narayan2020} proposes to employ compiler-based
techniques to force experts to write checks prior accessing interface-crossing
data (and in particular indexing information).  This ensures that humans will
sanitize the API, but does not offer correctness guarantees. Hardware memory
capabilities~\cite{Watson2015} address DC2 by offering bounds safety for C/C++,
with the limitations mentioned in DC1. Generally, bounds-checking
techniques~\cite{Szekeres2013} can address DC2.

\paragraph{DC3: Usage of Corrupted Object}

A victim compartment may use an object corrupted by a malicious compartment.
Examples include corrupted strings lacking \customtt{NULL} termination or
including arbitrary format string parameters, corrupted OS/libc constructs such
as \customtt{FILE*}, corrupted integers causing numeric
errors~\cite{CWE189}, etc. Corrupted objects may be control or
non-control data~\cite{Chen2005}. DC3 impact includes, in addition to DoS,
information leaks, or exposing read, write, or execute primitives. Vectors are
the same as DC1 and DC2. The fundamental difficulty to address DC3 is that the
validity of an interface-crossing object is entirely dictated by the semantics
of the API and of its users. The difficulty to extract this information
systematically is a well-known problem~\cite{Yun2016}.
RLBox~\cite{Narayan2020} partially addresses DC3 with automatic validity
checking and copy, for the types that allow it (e.g., strings), and forces
manual sanitization for others, with the drawbacks mentioned for DC1.  Even for
types that allow automatic sanitization such as C-style strings, checks remains
partial (checking \customtt{NULL}-termination, but not format string
parameters).  Hardware memory capabilities~\cite{Watson2015} address part of
the \emph{symptoms} of DC3 with full spatial memory safety, but cannot offer
complete protection: not all control and data attacks that could be mounted on
DC3 rely on spatial memory safety vulnerabilities.

\subsection{Cross-Compartment Temporal Violations (TV)}

\paragraph{TV1: Expectation of API Usage Ordering}

A victim compartment may expose functions (or callbacks) to other compartments
and assume call ordering, without enforcing it. For example, a compartment may
expose two functions \customtt{init()} and \customtt{work()}, expecting
\customtt{init()} $\rightarrow$ \customtt{work()}. Malicious compartments may
call \customtt{work()} first. The immediate impact can be any form of
undefined behavior in the victim, spatial or temporal depending on the context,
such as DoS, uninitialized pointer dereferences, use-after-frees,
synchronization bugs, etc. RLBox~\cite{Narayan2020} proposes tooling to enforce
callback ordering, but still requires manual detection and patching
of TV1.  This poses further difficulties when a compartment can be concurrently
queried. This could be coupled with API semantics inference techniques such as
APISan~\cite{Yun2016} to lighten the manual effort, but these techniques remain
incomplete. Generally, there is a need for more research in identifying and
enforcing compartment API usage ordering.

\paragraph{TV2: Usage of Corrupted Synchronization Primitive}

A victim compartment may use corrupted synchronization primitives (e.g., mutexes,
locks).  The impact includes, beyond DoS (deadlock), that of any race-condition
which could be leveraged to mount control-based attacks. TV2 vulnerabilities
are a special case of DC3 where the corrupted object is a synchronization
primitive.  Existing frameworks do not offer solutions to this class of
vulnerabilities: addressing TV2 is particularly challenging, as it requires
redesigning the way distrusting compartments cooperate in multithreaded contexts.

\paragraph{TV3: Shared-Memory Time-of-Check-to-Time-of-Use}

A victim compartment may check corrupted data in the shared memory. This may
allow a malicious compartment to corrupt the value after the check and before
the use, making the check useless. TV3 may lead to any previously mentioned
\CIV impact. TV3 vectors are shared objects with double fetches.  Existing
mitigations include forcing the copy of objects to a private region before
checking, as done by RLBox~\cite{Narayan2020} (with the genericity limitations
mentioned in DL1), or forbidding concurrent modification altogether as done by
Midas~\cite{Bhattacharyya2022} (that targets only kernel-space).

\subsection{Summary: \CIV Protections are in their Infancy}

No existing compartmentalization framework tackles all \CIV classes. Techniques
that can address a subset of \CIVs only offer partial and/or non-generic
solutions. Even the most comprehensive system, RLBox~\cite{Narayan2020}, relies
extensively on manual checking with no guarantees of check correctness. It is
likely that combining all the techniques required to achieve state-of-the-art
protection would result in an impractical performance overhead that contradicts
the initial motivation of using lightweight isolation technologies for
fine-grain compartmentalization.  This observation motivates our study
assessing the safety and complexity impact of \CIVs.

\section{\sys: Exploring CIVs with Fuzzing}
\label{sec:design-and-implementation}

\subsection{Assumptions and Threat Model}
\label{subsec:threat}

We assume an application decomposed into compartments that are mutually
distrusting. Compartments are defined as protected subsystems (as proposed by
Lampson~\cite{Lampson1974}), with private code, heap, and stack. Compartments
communicate through interfaces. If a pointer is passed through an interface,
the object it references is shared between the two compartments.  A protection
mechanism enforces spatial isolation: code cannot access private data or code
from other compartments. The compartmentalization framework enforces
cross-compartment control-flow integrity: one compartment can only call
explicit entry points exposed by other compartments. These assumptions fit the
vast majority of modern frameworks~\cite{Wu2012, VahldiekOberwagner2019,
Hedayati2019, Schrammel2020, Liu2017, Koning2017, Park2019, Bauer2021,
Sartakov2021, Lefeuvre2021, Lefeuvre2022, Agadakos2022}.


We assume a completely compromised compartment which we refer to as the
\emph{malicious} compartment: an attacker can execute arbitrary code in its
context. Compartments communicate through interfaces that respect the semantics
of function calls, with variable degrees of sanitization. The malicious
compartment attempts to misuse these interfaces to attack another compartment
called the \emph{victim}. The interface can be abused in either direction,
according to the caller/callee role of the malicious/victim compartments:

\paragraph{Safebox}

As the caller, the malicious compartment can abuse an interface \emph{exposed by the
victim} (callee). Vectors of corruption are function call arguments, data in
shared memory, and return values of callbacks invoked by the victim to be
executed in the context of the malicious compartment.  This corresponds to a
\emph{safebox} scenario, in which a trusted subsystem (e.g., libssl) is
protected from the rest of the system.

\paragraph{Sandbox}

As the callee, the malicious compartment can abuse an interface \emph{invoked
by the victim} (caller). Here, corruption vectors are return values, data in
shared memory, and parameters of callbacks invoked by the malicious compartment
to be executed in the victim's context. This is a \emph{sandbox} scenario, in
which an untrusted component (e.g., a 3rd-party library) is prevented from
accessing the rest of the system.

\subsection{Overview}
\label{subsec:overview}

To assess the impact of neglecting interface safety, we propose to fuzz
monolithic (non-compartmentalized) software at possible arbitrary
compartment boundaries, and analyze the set of \CIVs we uncover. \sys is an
\emph{in-memory} fuzzer~\cite{Sutton2007}: it instruments software targets
to hook into arbitrary interfaces, such as libraries, modules, functions, etc.

Because of their systematic nature, approaches based on static analysis may seem
enticing to explore interface-related issues. However, related works leveraging
such techniques fail to scale to more than simple programs~\cite{Hu2015}. Hence,
we take a pragmatic approach and rely on fuzzing. Our goal is further different
from existing in-memory/in-process/library/API fuzzers~\cite{Serebryany2016,
Sutton2007} because our tool needs to fuzz both ways (safebox/sandbox).
Hence,
we develop \sys from scratch.

\begin{figure}
\center
\includegraphics[width=0.87\linewidth]{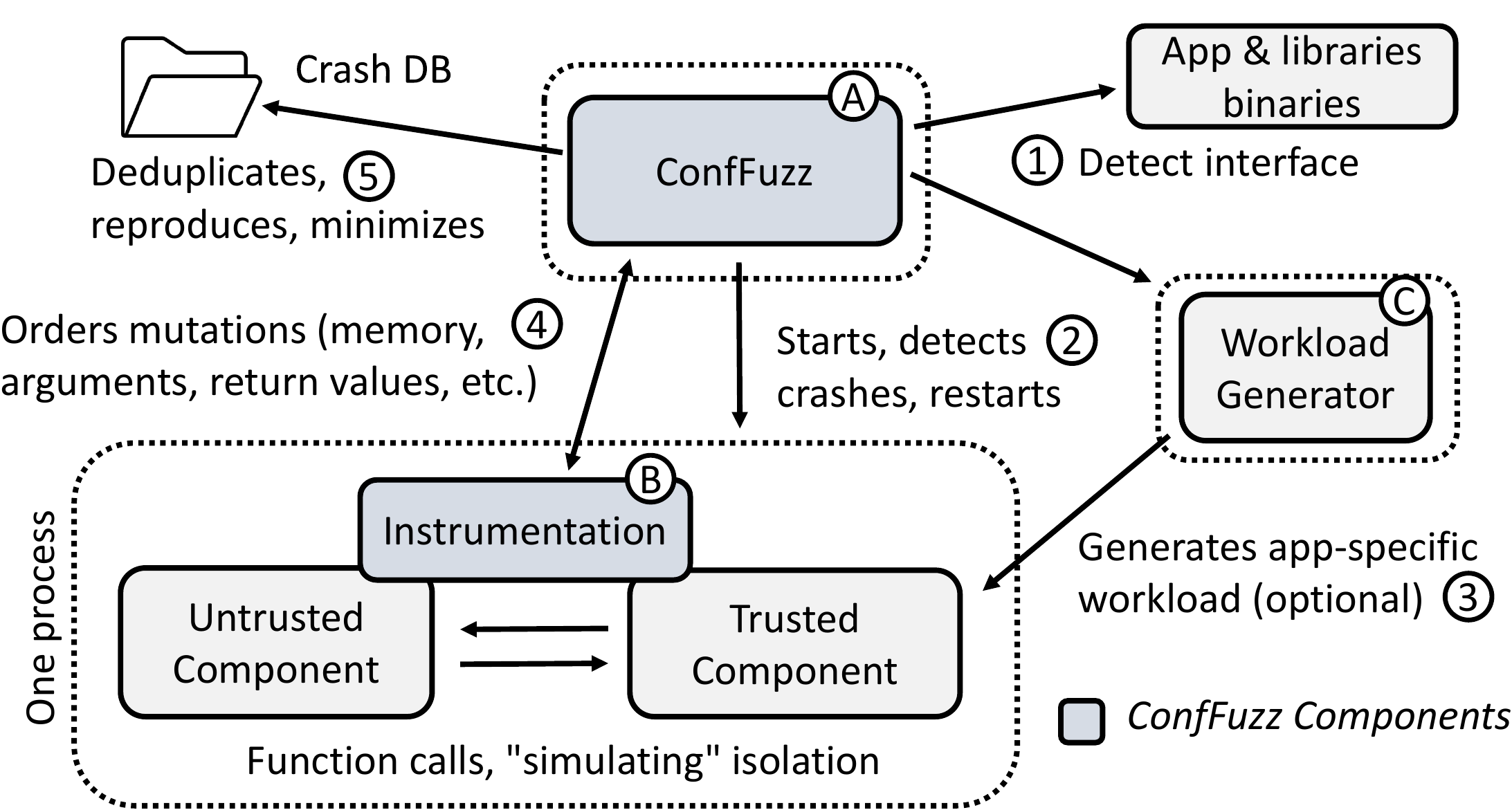}
\caption{
\sys architecture diagram.
}
\label{fig:architecture}
\vspace{-0.6cm}
\end{figure}

\paragraph{Two-Components Approach}
Each run of \sys considers two communicating software components: a malicious,
and a victim one. \sys simulates attacks towards the victim by automatically
altering data crossing the interface between them; we call this \emph{interface
data altering}.  For that, \sys hooks into the interface and fuzzes in
both directions (sandbox/safebox), altering function call arguments, shared
data, and return values for direct interface calls and callbacks.

\paragraph{Architecture Overview}
As shown in \Cref{fig:architecture}, \sys is composed of a self-contained
fuzzing monitor (\BC{A}), and dynamic binary instrumentation (DBI, \BC{B}).
This instrumentation, which sits between the malicious component and the victim
in the application's process, leverages the Intel Pin~\cite{INTELPIN} DBI
framework. Using Pin lets us apply \sys to software with a low engineering
cost, and hook at arbitrary interfaces unlike other approaches e.g.,
\customtt{LD\_PRELOAD}.

First, \sys automatically identifies the interface between application and
compartment components by analyzing debug information in the corresponding
binaries (\BC{1}). When starting the application (\BC{2}), the fuzzing monitor
dynamically injects instrumentation wrappers at the detected compartment
interface. Following this, it optionally starts a workload generator (\BC{C})
to stimulate the application (\BC{3}). At runtime, at each API call, the
fuzzing logic in the monitor determines a set of alterations to perform,
possibly through mutation of an existing set, and performs the alterations via
the instrumentation (\BC{4}).

The application runs with Address Sanitizer~\cite{Serebryany2012} (ASan) as bug
detector. When ASan reports a crash in the victim, \sys deduplicates it based
on the stack trace. If the bug is not known, \sys reproduces it, before
minimizing the set of alterations performed to obtain the nucleus of
alterations that trigger the bug (\BC{5}). \sys is implemented in 4.5K~LoC of
C++, Bash and Python. The following subsections present \sys's fuzzing process
steps in greater details.

\subsection{Interface Detection and Instrumentation}

\sys automatically handles the detection of the signature of a given target
API. The tool gathers a list of API symbols: functions and callbacks used by
the victim and malicious components to communicate with each other, along with,
for each of these, the number of arguments, the type and size of each argument,
and the type and size of the return values if appropriate. We compile the
target application with debug symbols, allowing \sys to use DWARF metadata to
retrieve interface and type information. The tool can automatically infer the
list of functions composing the interface exposed by shared libraries, and the
user can provide that list manually when targeting arbitrary interfaces. Most
functions are instrumented when the application starts. Concerning callbacks,
\sys automatically detects them at runtime by scanning API call parameters for
function pointers, and instrumenting the identified functions on the fly. \sys
automatically infers what data is shared between the malicious and victim
components, considering that all buffers referenced by pointers crossing the
API are shared data.

Each symbol, including API elements and callbacks, is instrumented at entry and
exit. At each of these events, the instrumentation checks for reentrance to
protect against API calls performed by the compartmentalized component itself,
notifies the fuzzing monitor with information about the function being executed
and arguments/return value information, and allows the monitor to perform
alterations: depending on the type of symbol and the fuzzing direction
(safe/sandbox), altering argument values, return value, altering shared memory,
etc. The instrumentation is kept as simple as possible and all the fuzzing
logic runs in the monitor.  Instrumentation and monitor communicate via a
well-defined protocol using pipes.

\subsection{Workload Generation and Coverage}

\sys passively sits at internal API boundaries in an application, and users
must determine an application-specific configuration and input workload that
exercise the API. Finding a good set of inputs with high coverage is a problem
shared across most fuzzers~\cite{Rebert2014, Zhang2022}. In the data set
considered in this paper, the time to understand the configuration system of an
application and find an appropriate workload went from a few minutes to a few
hours for a graduate student. We also explored the use of
OSS-Fuzz~\cite{OSS_FUZZ} to generate workloads for the application, but found
that hand-tuned workloads are generally better at precisely targeting the
internal APIs that we intend to fuzz, which is critical for this study.
Nevertheless, OSS-Fuzz should be considered in cases where the manual effort to
create workloads should be minimized.

\sys is not coverage-guided. However, to provide an indication of how
comprehensive the fuzzing of an API is with a given configuration and workload,
\sys measures \emph{API coverage}: the number of target API functions reached.
This metric can be compared to the target API's size to understand the
coverage of a given configuration and workload.

\subsection{Interface Data Altering and Fuzzing Strategy}

\begin{table}
\centering
\caption{\emph{\sys data altering strategies for each CIV class}. Each class of data alterations done by \sys is targeted at a particular type of \CIV, as shown in this table.}
{
\setlength{\tabcolsep}{5pt}
\footnotesize
\begin{tabular}{|c|p{6.5cm}|}
\hline
\CIV Class & Corresponding Data Alteration Strategy \\ \hline \hline
\multirow{2}{*}{DC1} & \emph{Alteration of pointer types} to invalid values (zero page, arbitrary unmapped areas). \\ \hline
\multirow{5}{*}{DC2} & \emph{Alteration of integer types}: increments/decrements to trigger over/underflows, replacement to known limits such as \tinytt{INT\_MAX} (possibly at offsets) to trigger numeric errors, replacement with random values. \\ \hline
DC3 & \parbox{6.5cm}{\emph{Alteration of non-pointer types \& of pointer targets}: decrements/increments at varying offsets in the object, replacement of bytes at various offsets in the object.\\\emph{Replacement of pointers to the same type} (replay), and \emph{Replacement of pointers to different types} (type confusion) to trigger more complex DC3 flaws.} \\ \hline
\multirow{1}{*}{TV1} & \emph{Non execution of API functions} for partial TV1 detection. \\ \hline
\multirow{2}{*}{TV2} & \emph{Alteration of mutex/lock types} in shared memory (allows partial detection of TV2). \\ \hline
TV3, DL1-2 & Not targeted. \\ \hline
\end{tabular}
}
\label{table:alteringstrategies}
\end{table}

At each API crossing, the instrumentation notifies the monitor, which may
proceed to alter interface data over the entire attack surface exposed to the
malicious component. When fuzzing in sandbox mode, the malicious component
may alter return values and callback arguments. In safebox mode, function
call arguments and callback return values may be altered. In both cases, the
malicious component may also alter data shared between the malicious and
victim components.

\sys probabilistically decides whether or not to alter data at an API crossing.
In order to avoid revealing only shallow bugs that systematically crash in
early fuzzing stages, we use a dynamic probability adaptation threshold: at
first, the threshold is at 0, i.e., \sys alters data aggressively at all
crossings. When the number of new crashes becomes scarce, \sys increments the
threshold to find crashes further in the API usage. Concretely, based on a
counter incremented at each API crossing, crossings that come before the
threshold is reached see their data altered with a lower probability. This
allows \sys to find stateful crashes as well.

\sys alters values in two ways: applying increments/decrements, and replacing
the value altogether. \sys uses type information to drive alterations. For
pointer values, \sys may perform replacements with other pointers of the same
type, of different types, at varying offsets, at the zero page, on the heap,
stack, data, text sections, etc. For integer values, \sys may perform
replacements with known limits (e.g., \customtt{INT\_MAX}). A detailed
description of data alterations performed by \sys is provided in
\Cref{table:alteringstrategies}.  While fuzzing, the fuzzer enriches an
alteration corpus with values gathered during previous alterations. Values from
the corpus are reused, possibly mutated, with a given probability.

\subsection{Crash Processing and Bug Analysis}

\paragraph{Crash Sanitization}

Upon a crash, \sys compares the ASan stack trace with its database of known
crashes. If the crash is a duplicate, no further analysis is performed, but
information about the new occurrence is logged. \sys then checks whether the
new crash is a false positive. False positives arise when altered objects are
sent back to the malicious component by the victim. In such cases the malicious
component corrupts itself, yielding an invalid bug.

In order to detect false positives, \sys walks down the stack trace until it finds an
entry referencing code belonging to a component. If that component is the
one considered as malicious for this fuzzing run, the crash is considered a
false positive.  Even though such false positives are not valid crashes, \sys
still attempts to minimize them, as this allows to detect non-viable data
alterations that can be avoided later in order to minimize time wasted on false
positives.

\paragraph{Reproduction and Minimization}

After sanitization, \sys systematically attempts to reproduce crashes.
Unfortunately, not all crashes are reproducible, as some might be due to
particular non-deterministic factors such as scheduling effects, reliance by the
application on random values/changing external inputs, etc. A non-reproducible
crash cannot be further processed automatically by \sys. Still, information
regarding such crashes are valuable for the analysis and are logged for manual
inspection. On the other hand, if the crash can be reproduced, the monitor
gradually minimizes it.

As part of the minimization step, \sys tries to understand the minimum set of
alteration steps required to trigger a given bug. \sys gradually goes through
each alteration performed in reverse order (since the last alterations
performed tend to be the most likely to trigger the crash), and determines
whether the alteration is sufficient to trigger the crash, necessary to trigger
the crash (without it the crash cannot be reproduced, but it is not sufficient
by itself), or superfluous. This results in a minimal set of attack primitives
that the malicious component can perform to trigger the bug.

\paragraph{Impact Analysis of Crashes}

After processing crashes, \sys performs initial triage. It harvests
ASan-provided information: whether the crash is due to an illegal read, write
or execution, allocator corruption, or to a \customtt{NULL} dereference, along
with faulty addresses. Then, for R/W/X crashes, \sys tries to determine whether
the vulnerability is arbitrary: for each alteration in the minimized steps,
\sys mutates the altered value with increments/decrements, trying to reproduce
the crash. If the crash is reproducible and the faulty address varies
accordingly to the increment/decrement, the vulnerability is considered
arbitrary.

\section{A Large-Scale Study of Real-World \CIVs}\label{sec:evaluation}

\begin{table*}
\centering
    \caption{Bugs found for sandbox and safebox Trust Models (TM,
        fuzzing directions). \emph{Refs.} links to studies that implemented an
        equivalent scenario. \emph{Victims} gives the number of individual
        software components (application code, libraries, modules) that the
        malicious component managed to crash. \emph{API Coverage} represents
        workload coverage: \emph{callers} denotes the number of components calling
	the API, and \emph{coverage} denotes how many functions of the fuzzed API are hit
        at runtime. \emph{Impact} describes the type of bug: R/W/X
        fault, \customtt{NULL} dereference, or improper calls to the allocator
        (e.g. calling \customtt{malloc} with a negative value).}

{
\setlength{\tabcolsep}{5pt}

\scriptsize
\begin{tabular}{|c|c|l|c|l|l|c|c|c|l|l|l|l|l|}
\hline
\multicolumn{1}{|c|}{\multirow{2}{*}{\textbf{TM}}} & \multicolumn{1}{|c|}{\multirow{2}{*}{\textbf{Application}}} & \multicolumn{1}{|c|}{\multirow{2}{*}{\textbf{Compartment API}}} & \multicolumn{1}{|c|}{\multirow{2}{*}{\textbf{References}}} & \multicolumn{2}{|c|}{\multirow{1}{*}{\textbf{Crashes}}} & \multirow{2}{*}{\textbf{Victims}} & \multicolumn{2}{|c|}{\multirow{1}{*}{\textbf{API Coverage}}} & \multicolumn{5}{|c|}{\textbf{Impact} \footnotesize{(of which arbitrary)}} \\ \cline{5-6}\cline{8-14}
\multicolumn{1}{|l|}{} & & & & \textit{Raw} & \textit{Dedup.} & & \textit{Callers} & \textit{Coverage} & \textit{Read} & \textit{Write} & \textit{Exec} & \textit{Alloc} & \textit{Null} \\ \hline \hline
\multicolumn{1}{|c|}{\multirow{31}{*}{\rotatebox{90}{\emph{Sandbox}}}} & \multicolumn{1}{|l|}{\multirow{2}{*}{HTTPd}} & libmarkdown    & \cite{Narayan2020}       & 192 & 13 & 3 & 1 & 100\% (4/4)  & 10 (8) & 7 (7)  & 0 (0)& 1 & 4  \\ \cline{3-14}
\multicolumn{1}{|l|}{} & \multicolumn{1}{|l|}{}                                                                       & mod\_markdown  &                          & 381 & 71 & 5 & 1 & 100\% (1/1)  & 62 (52)& 17 (14)& 2 (1)& 0 & 30 \\ \cline{2-14}
\multicolumn{1}{|l|}{} & \multicolumn{1}{|l|}{aspell}                                                                 & libaspell      &                          & 278 & 8  & 1 & 1 & 34\% (48/141)& 7 (7)  & 7 (7)  & 2 (1)& 0 & 3  \\ \cline{2-14}
\multicolumn{1}{|l|}{} & \multicolumn{1}{|l|}{bind9}                                                                  & libxml2 (write API) &                     & 0   & 0  & 0 & 1 & 86\% (13/15) & 0 (0)  & 0 (0)  & 0 (0)& 0 & 0  \\ \cline{2-14}
\multicolumn{1}{|l|}{} & \multicolumn{1}{|l|}{bzip2}                                                                  & libbz2         & \cite{Wu2012, Bauer2021} & 16  & 5  & 1 & 1 & 62\% (5/8)   & 5 (2)  & 1 (0)  & 0 (0)& 0 & 0  \\ \cline{2-14}
\multicolumn{1}{|l|}{} & \multicolumn{1}{|l|}{cURL}                                                                   & libnghttp2     &                          & 61  & 7  & 2 & 1 & 50\% (18/36) & 3 (3)  & 5 (5)  & 0 (0)& 1 & 3  \\ \cline{2-14}
\multicolumn{1}{|l|}{} & \multicolumn{1}{|l|}{exif}                                                                   & libexif        &                          & 400 & 7  & 1 & 1 & 10\% (13/129)& 3 (3)  & 0 (0)  & 0 (0)& 0 & 5  \\ \cline{2-14}
\multicolumn{1}{|l|}{} & \multicolumn{1}{|l|}{\multirow{3}{*}{FFmpeg}}                                                & libavcodec     &                          & 316 & 20 & 3 & 4 & 31\% (19/60) & 13 (12)& 12 (12)& 0 (0)& 3 & 7  \\ \cline{3-14}
\multicolumn{1}{|l|}{} & \multicolumn{1}{|l|}{}                                                                       & libavfilter    &                          & 51  & 1  & 1 & 2 & 12\% (2/16)  & 1 (1)  & 0 (0)  & 0 (0)& 0 & 1  \\ \cline{3-14}
\multicolumn{1}{|l|}{} & \multicolumn{1}{|l|}{}                                                                       & libavformat    &                          & 217 & 9  & 2 & 3 & 52\% (10/19) & 8 (7)  & 1 (1)  & 0 (0)& 0 & 7  \\ \cline{2-14}
\multicolumn{1}{|l|}{} & \multicolumn{1}{|l|}{file}                                                                   & libmagic       &                          & 150 & 5  & 1 & 1 & 63\% (7/11)  & 5 (2)  & 1 (1)  & 0 (0)& 0 & 4  \\ \cline{2-14}
\multicolumn{1}{|l|}{} & \multicolumn{1}{|l|}{\multirow{2}{*}{git}}                                                   & libcurl        & \cite{Im2021}            & 13  & 4  & 2 & 1 & 90\% (18/20) & 2 (2)  & 2 (2)  & 0 (0)& 1 & 1  \\ \cline{3-14}
\multicolumn{1}{|l|}{} & \multicolumn{1}{|l|}{}                                                                       & libpcre        &                          & 81  & 2  & 1 & 1 & 44\% (8/18)  & 2 (2)  & 0 (0)  & 0 (0)& 2 & 0  \\ \cline{2-14}
\multicolumn{1}{|l|}{} & \multicolumn{1}{|l|}{\multirow{2}{*}{Inkscape}}                                              & libpng         & \cite{Wu2012}            & 66  & 3  & 1 & 1 & 46\% (14/30) & 2 (1)  & 2 (2)  & 0 (0)& 0 & 1  \\ \cline{3-14}
\multicolumn{1}{|l|}{} & \multicolumn{1}{|l|}{}                                                                       & libpoppler     & \cite{Gudka2015}         & 81  & 4  & 2 & 1 & 100\% (9/9)  & 4 (3)  & 4 (4)  & 0 (0)& 0 & 2  \\ \cline{2-14}
\multicolumn{1}{|l|}{} & \multicolumn{1}{|l|}{libxml2-tests}                                                          & libxml2 (write API) &                     & 0   & 0  & 0 & 1 & 100\% (47/47)& 0 (0)  & 0 (0)  & 0 (0)& 0 & 0  \\ \cline{2-14}
\multicolumn{1}{|l|}{} & \multicolumn{1}{|l|}{lighttpd}                                                               & mod\_deflate   &                          & 117 & 26 & 2 & 1 & 100\% (6/6)  & 16 (11)& 5 (0)  & 1 (1)& 2 & 9  \\ \cline{2-14}
\multicolumn{1}{|l|}{} & \multicolumn{1}{|l|}{\multirow{3}{*}{\parbox{0.2cm}{Image\\Magick}}}                         & libghostscript & \cite{Bauer2021}         & 67  & 14 & 2 & 1 & 100\% (11/11)& 4 (2)  & 1 (1)  & 0 (0)& 3 & 9  \\ \cline{3-14}
\multicolumn{1}{|l|}{} & \multicolumn{1}{|l|}{}                                                                       & libpng         & \cite{Wu2012}            & 778 & 44 & 1 & 2 & 22\% (17/77) & 2 (2)  & 9 (9)  & 2 (0)& 2 & 39 \\ \cline{3-14}
\multicolumn{1}{|l|}{} & \multicolumn{1}{|l|}{}                                                                       & libtiff        & \cite{Wu2012}            & 197 & 14 & 2 & 1 & 30\% (13/43) & 3 (3)  & 6 (6)  & 0 (0)& 0 & 13 \\ \cline{2-14}
\multicolumn{1}{|l|}{} & \multicolumn{1}{|l|}{\multirow{2}{*}{Nginx}}                                                 & libpcre        &                          & 144 & 10 & 1 & 1 & 93\% (14/15) & 8 (7)  & 3 (3)  & 0 (0)& 6 & 2  \\ \cline{3-14}
\multicolumn{1}{|l|}{} & \multicolumn{1}{|l|}{}                                                                       & mod\_geoip     & \cite{Schrammel2022}     & 276 & 25 & 2 & 1 & 35\% (5/14)  & 21 (17)& 4 (1)  & 1 (1)& 1 & 10 \\ \cline{2-14}
\multicolumn{1}{|l|}{} & \multicolumn{1}{|l|}{\multirow{2}{*}{Okular}}                                                & libmarkdown    & \cite{Narayan2020}       & 64  & 5  & 3 & 1 & 100\% (4/4)  & 3 (1)  & 0 (0)  & 0 (0)& 1 & 2  \\ \cline{3-14}
\multicolumn{1}{|l|}{} & \multicolumn{1}{|l|}{}                                                                       & libpoppler     & \cite{Gudka2015}         & 195 & 9  & 1 & 1 & 6\% (24/379) & 8 (6)  & 7 (7)  & 0 (0)& 1 & 4  \\ \cline{2-14}
\multicolumn{1}{|l|}{} & \multicolumn{1}{|l|}{\multirow{2}{*}{Redis}}                                                 & mod\_redisbloom&                          & 389 & 23 & 1 & 1 & 42\% (8/19)  & 18 (13)& 6 (4)  & 0 (0)& 0 & 13 \\ \cline{3-14}
\multicolumn{1}{|l|}{} & \multicolumn{1}{|l|}{}                                                                       & mod\_redisearch&                          & 381 & 21 & 1 & 1 & 54\% (18/33) & 15 (14)& 14 (11)& 0 (0)& 0 & 12 \\ \cline{2-14}
\multicolumn{1}{|l|}{} & \multicolumn{1}{|l|}{rsync}                                                                  & libpopt        &                          & 167 & 8  & 1 & 1 & 90\% (9/10)  & 4 (3)  & 2 (0)  & 0 (0)& 0 & 5  \\ \cline{2-14}
\multicolumn{1}{|l|}{} & \multicolumn{1}{|l|}{squid}                                                                  & libxml2        &                          & 226 & 12 & 1 & 1 & 70\% (7/10)  & 9 (5)  & 3 (3)  & 4 (1)& 0 & 4  \\ \cline{2-14}
\multicolumn{1}{|l|}{} & \multicolumn{1}{|l|}{su}                                                                     & libaudit       &                          & 0   & 0  & 0 & 1 & 66\% (2/3)   & 0 (0)  & 0 (0)  & 0 (0)& 0 & 0  \\ \cline{2-14}
\multicolumn{1}{|l|}{} & \multicolumn{1}{|l|}{\multirow{2}{*}{Wireshark}}                                             & libpcap        &                          & 162 & 8  & 2 & 1 & 50\% (20/40) & 8 (3)  & 5 (5)  & 0 (0)& 0 & 4  \\ \cline{3-14}
\multicolumn{1}{|l|}{} & \multicolumn{1}{|l|}{}                                                                       & libzlib        &                          & 42  & 1  & 1 & 1 & 85\% (6/7)   & 0 (0)  & 0 (0)  & 0 (0)& 0 & 1  \\ \hline \hline
\multicolumn{1}{|l|}{} & \multicolumn{3}{|l|}{Total:} & 5508 & 379 & 47 & 38 & N/A & 246 (192) & 124 (105) & 12 (5) & 24 & 195 \\ \hline\hline
\multicolumn{1}{|c|}{\multirow{8}{*}{\rotatebox{90}{\emph{Safebox}}}} & \multicolumn{1}{|l|}{cURL}         & libssl               & \cite{Bauer2021}              & 198 & 27 & 1 & 1 & 25\% (14/56) & 18 (10) & 5 (4)  & 1 (1) & 0 & 17 \\ \cline{2-14}
\multicolumn{1}{|l|}{} & \multicolumn{1}{|l|}{GPA}                                                         & libgpgme             &                               & 174 & 9  & 1 & 1 & 4\% (3/72)   & 7 (2)   & 0 (0)  & 0 (0) & 0 & 6  \\ \cline{2-14}
\multicolumn{1}{|l|}{} & \multicolumn{1}{|l|}{GPG}                                                         & libgcrypt            & \cite{Bauer2021}              & 4221& 105& 1 & 1 & 15\% (15/95) & 64 (60) & 4 (0)  & 0 (0) & 77& 20 \\ \cline{2-14}
\multicolumn{1}{|l|}{} & \multicolumn{1}{|l|}{Memcached}                                                   & internal\_hashtable  & \cite{Park2019}               & 4037  & 16 & 1 & 1 & 50\% (6/12)  & 10 (5)  & 2 (0)  & 0 (0) & 1 & 6  \\ \cline{2-14}
\multicolumn{1}{|l|}{} & \multicolumn{1}{|l|}{\multirow{2}{*}{Nginx}}        & internal\_libssl-keys& \cite{Park2019, VahldiekOberwagner2019, Gu2022, Litton2016} & 599 & 46 & 1 & 1 & 50\% (2/4)   & 32 (1)  & 28 (0) & 0 (0) & 0 & 22 \\ \cline{3-14}
\multicolumn{1}{|l|}{} & \multicolumn{1}{|l|}{}                                                  & libssl  & \cite{Bauer2021, Agadakos2022, Im2021, Sartakov2021} & 346 & 39 & 2 & 1 & 11\% (11/96) & 16 (13) & 19 (13)& 2 (1) & 0 & 26 \\ \cline{2-14}
\multicolumn{1}{|l|}{} & \multicolumn{1}{|l|}{\multirow{2}{*}{sudo}}                                       & internal\_auth-api   &                               & 191 & 5  & 1 & 1 & 100\% (5/5)  & 5 (4)   & 0 (0)  & 0 (0) & 0 & 4  \\ \cline{3-14}
\multicolumn{1}{|l|}{} & \multicolumn{1}{|l|}{}                                                            & libapparmor          &                               & 97  & 3  & 1 & 1 & 100\% (2/2)  & 2 (2)   & 2 (0)  & 0 (0) & 0 & 2  \\ \hline \hline
\multicolumn{1}{|l|}{} & \multicolumn{3}{|l|}{Total:} & 9863 & 250 & 9 & 8 & N/A & 154 (97) & 60 (17) & 3 (2) & 78 & 103\\ \hline
\end{tabular}
\label{tab:fuzzingtable}
}
\end{table*}

In this section, we use \sys to gather a large dataset of real-world \CIVs, from
which we extract insights on a set of research questions: (Q1) what is the
number of CIVs at legacy, unported APIs?, (Q2) what patterns lead to \CIVs and
are all APIs similarly affected by CIVs?, (Q3) what is the complexity to address
these CIVs when compartmentalizing?, and (Q4) what is the range of severity of
the CIVs we uncover, i.e. without a fix, what can attackers do?
\Cref{tab:fuzzingtable} shows our results. \S\ref{subsec:methodology} and
\S\ref{subsec:datasetoverview} give an overview of the methodology and results.
The following sections provide in-depth analysis.

\subsection{Methodology}
\label{subsec:methodology}

\paragraph{Choice of Scenarios}

Our corpus is composed of \numberofapps applications and \numberoflibs
library/module/function APIs, totaling \numberofscenarios real-world
sandbox and safebox scenarios.  We choose scenarios that are meaningful
security-wise, e.g., sandboxing image processing libraries because they are
higher-risk, or safeboxing functions manipulating SSL keys because they are
sensitive.  Motivated by (Q2), we choose scenarios that encompass a diversity
of API types (libraries, pluggable modules, internal APIs), and usages (image
processing, text parsing, logging, key management, etc). We focus on highly
popular applications. Several of these scenarios
(\numberofscenariosfromliterature/\numberofscenarios, see table) have been
presented as use cases in \numberofframeworks different compartmentalization
frameworks, only \numberofframeworksthatconsidercivs of
which~\cite{Narayan2020, Gudka2015} providing protection against certain \CIV
classes (discussed in \S\ref{sec:taxonomy}).  The smaller number of safebox
scenarios reflects that, in general, libraries/modules tend to perform
more untrusted operations that one might want to sandbox (e.g., request
processing, complex data parsing) than trusted operations that one might want
to safebox (e.g., cryptographic operations).

\paragraph{Comparison with Related Works}

As discussed in \S\ref{subsec:overview}, existing fuzzers are unfit by design
to the investigation of \CIVs, so we present no baseline. A relevant work in
the domain of static analysis is DUI Detector~\cite{Hu2015}, which could be
used to detect classes DC1 and DC2 presented in \S\ref{par:DUIs}.
Unfortunately, its authors were unable to provide us with its source code. We
therefore provide a textual comparison in \S\ref{sec:related-works}. We
evaluated the cost of DBI (Pin) in ConfFuzz on a representative set of
applications (Nginx, Redis, file, xmltest), and found it to be $\le$65\%,
thanks to Pin's probe (JIT-less) instrumentation mode. These numbers match
official data~\cite{Devor2013}. The overhead could be further reduced
with compiler-inserted instrumentation, which we leave for future works.

\subsection{Overview of the Data Set}
\label{subsec:datasetoverview}

Overall, \sys found \numberofcrashes unique bugs deduplicated from
\numberofcrashespriordeduplication crashes.  \sys uncovered bugs of 4 different
classes as listed in \S\ref{sec:taxonomy}: DC1-3, and TV2.  In the
\emph{sandbox} mode, where \sys fuzzes from the (untrusted) component towards
components calling it (e.g., application code, other libraries/modules), we
found \numberofsandboxcrashes \CIVs. For \numberofscenarioswithoutcrashes
scenarios, \sys did not find any \CIVs. We discuss these scenarios in
\S\ref{subsec:patterns}.  In the \emph{safebox} mode, the fuzzer found
\numberofsafeboxcrashes unique crashes. There are significant differences with
the sandbox results impact distribution, which we analyze in
\S\ref{subsec:patterns}.  For both modes, the summed number of bugs of all
impact types is larger than the total bug count (\sumofimpacts versus
\numberofcrashes) because \numberofbugswithmultipleimpact bugs (1/3rd of the
bugs) have more than one type of impact.

\paragraph{API End-Point Coverage}

We observe unequal interface coverage across scenarios. Some scenarios such as
HTTPd with libmarkdown show full API coverage, while others such as bzip2 with
libbz2 feature poorer coverage.  Generally, low coverage is due to \sys fuzzing
a single software configuration or CLI option. In the case of bzip2 for
example, \sys fuzzes decompression (\customtt{-d}), but not
compression or archive testing (\customtt{-z/-t}).  These entry points of libbz2
are left unexplored, and the user is notified.  This is a common problem across
fuzzers, addressable by varying configurations~\cite{Rebert2014, Zhang2022},
and fuzzing with more workloads, which we consider out of scope.  Nevertheless,
despite its simple exploration approach, \sys shows good ability to find
relevant bugs. For example, in HTTPd with libmarkdown, a scenario from
RLBox~\cite{Narayan2020}, \sys correctly discovered all \CIVs addressed by
RLBox, asserted the criticality of the bugs, and even one more due to library
version differences.

\subsection{Prevalence of \CIVs}
\label{subsec:prevalence}

\begin{figure*}
    \center
    \includegraphics[width=1.0\textwidth]{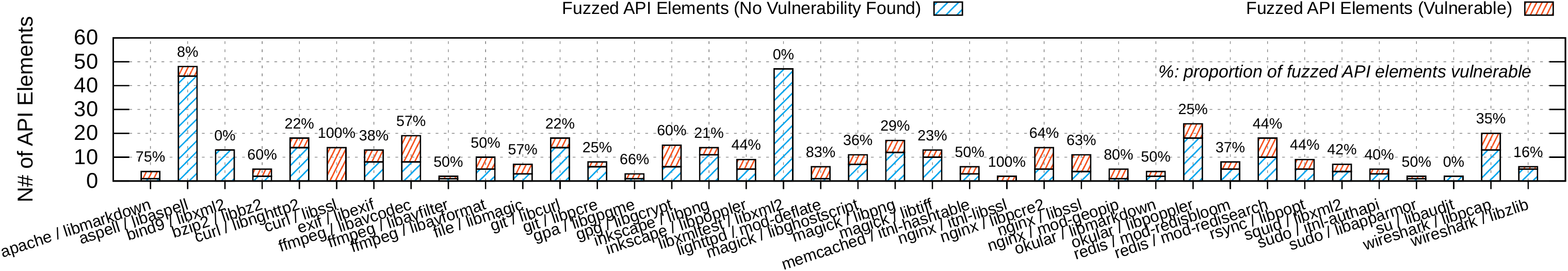}
    \caption{Proportion of covered vulnerable endpoints versus covered
	     endpoints for each scenario (see \Cref{tab:fuzzingtable} for
	     coverage).}
    \label{fig:apivuln}
    \vspace{-0.2cm}
\end{figure*}

At the highest level, the results confirm our expectations: \emph{\CIVs are
widespread among unmodified applications}. Indeed, all but 3 scenarios present
\CIVs. Looking into greater details, however, disparities appear: libraries
present widely varying \CIV numbers, ranging 0-105 for a single scenario.
Disparities become even clearer when looking at the ratio of vulnerable over
covered API elements, ranging 0\%-100\%, as shown in \Cref{fig:apivuln}.  This
observation is not a consequence of coverage disparities; we find that the
number of functions covered and the number of \CIVs found are uncorrelated
($|r| < 0.09$). Similarly, there is no correlation between the size of
compartment APIs and the number of \CIVs found ($|r| < 0.1$). This observation
further materializes when considering APIs that do comparable tasks, e.g.,
libbz2 and libzlib (compression), or libpng and libtiff in ImageMagick
(graphics).  In both cases, \CIV counts vary by 3-5x for nearly identical
coverage.  This is most startling in ImageMagick where the workload (image
format conversion) is identical in both cases, but the number of \CIVs jumps
from 14 to 44. These observations hint that the vulnerability of interfaces to
\CIVs is a factor of \emph{individual patterns and structural properties}
rather than of size or functionalities. We study these patterns in the next
section.

\insight{
  smaller APIs do not imply less \CIVs in unmodified
  applications: API size and number of \CIVs are uncorrelated.
}

\subsection{Patterns Leading to the Presence/Absence of \CIVs}
\label{subsec:patterns}

\begin{figure*}
    \center
    \includegraphics[width=1.0\textwidth]{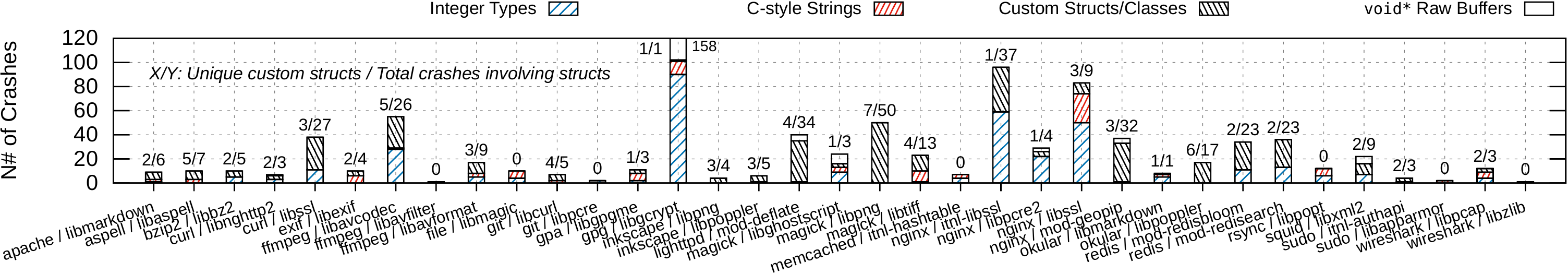}
    \caption{High-level type classes involved in \CIVs for each scenario.}
    \label{fig:typesvuln}
    \vspace{-0.3cm}
\end{figure*}

Zooming in, the types/data structures triggering crashes are integer types, custom
classes and structures, C-style strings, and raw \customtt{void*} buffers flowing
through compartment interfaces.  As shown in \Cref{fig:typesvuln}, integer
types and custom objects are the main culprits, being involved in 85\% of
\CIVs, while only 33\% of \CIVs are due to C-style strings and raw buffers.
\Cref{table:lowlevelpatterns} summarizes the low-level \CIV patterns observed
for each type. The following paragraphs give insights from a selection of
these patterns and higher-level observations.

\begin{table}[]
\centering
\caption{Low-level \CIV patterns from API-crossing types.}
{
\setlength{\tabcolsep}{5pt}
\scriptsize
\begin{tabular}{|l|p{6.5cm}|}
\hline
Type Class             & Low-Level \CIV Patterns Involving this Type Class \\ \hline \hline
\multirow{2}{*}{\emph{Integer Types}}   & sizes and bounds (DC2), error codes and return values (DC3, Pattern 4), and generally control-flow manipulations (DC3) \\ \hline
\multirow{2}{*}{\emph{Custom Structs}}  & state data (DC3, Pattern 1), mutexes (TV2, Pattern 1), non-control data (DC3) \\ \hline
\emph{C-style Strings} & version/error strings (Pattern 4), serialized data (all DC1\&3) \\ \hline
\multirow{2}{*}{\emph{void* buffers}}   & exposed allocator arguments (DC1-3, Pattern 5), and generally opaque data structures (DC1\&3, rare in this dataset) \\ \hline
\end{tabular}
}
\label{table:lowlevelpatterns}
\end{table}

\paragraph{Pattern 1 -- Modularity \& Exposure of Internal State}


Module APIs are some of the most vulnerable interfaces in the study; HTTPd,
Nginx, Redis, and lighttpd modules rank 2nd-10th in \CIV count, clearly above
average. HTTPd, particularly, ranks 2nd with 71 unique \CIVs, even though its
module API has a single function and entry point. Impact-wise, modules
represent more than half of read \CIVs, and 1/3rd of write and execute \CIVs,
even though they only represent 5/\numberofscenarios of our scenarios. In
short, the most modular interfaces of the dataset get \emph{more bugs} and
\emph{worse bugs} on average.

We track down these observations to one common pattern across module APIs:
\emph{exposing the application's internals to maximize performance and
flexibility}.  Unlike library APIs, module APIs are designed to accommodate
\emph{generic needs with good performance}.  These needs stand at odds with the
requirements of compartmentalization: in order to achieve this, the application
must expose its internals to the module, resulting in significantly less
encapsulation than with traditional libraries.  Take Apache modules as an
example: HTTPd modules can register hooks into core operations of the server
(e.g., configuration, name translation, request processing), and are
systematically passed a request structure \customtt{request\_rec}. This
structure is highly complex, with over 75 fields, of which over 60\% of
pointers, many of them referencing other complex structures: memory pools,
connection data, server data, and even synchronization structures (resulting in
TV2 \CIVs). These shared structures are not only very hard to sanitize, they
lead to oversharing when compartmentalizing because most modules do not need
access to all of them. These observations become even more clear comparing the
results of the two Apache scenarios: mod\_markdown and libmarkdown. Here, the
HTTPd Markdown module uses libmarkdown under the hood, thus we can isolate at
the module boundary, or at the library boundary to obtain comparable
guarantees.  As expected, isolating at the library boundary yields fewer
crashes than isolating at the module boundary (5.5x less). Despite a larger
number of entry points, the attack surface exposed to the library is far
smaller than that of the module API due to encapsulation; the library does not
have knowledge or access to Redis' internals.  We make similar observations for
Nginx, Redis, and lighttpd.  These observations are not trivial: multiple
studies propose privilege separation of modules~\cite{Im2021, Schrammel2022}
without considering this problem, and other studies~\cite{Li2021, Sartakov2021,
Lefeuvre2022} build on the assumption of modularity being fundamentally good
for compartmentalization.  We stress that this is not always the case, and that
more research is need to achieve fast, generic, and compartmentalizable modular
APIs.

\insight{
  modularity $\neq$ low compartmentalization complexity:
  boundaries exposing internal state are hard to protect.
}

\paragraph{Pattern 2 -- \CIV-Resilient Input-Only APIs}

Several scenarios are entirely \CIV-free: libxml2 write API, and su with libaudit.  In
the case of libxml2, even API tests with full coverage of all 47 endpoints do
not yield a single \CIV. We manually confirmed that no \CIVs are possible at
these APIs, reducing these observations to one common pattern:
\emph{designing the API like a write/input-only endpoint}. For instance, Bind9
uses libxml2 to store statistics onto the filesystem in XML format; it strictly
forwards data to libxml2, which formats and stores it. The situation is similar
in su with libaudit, used as a logging facility. su passes logs to libaudit,
which processes them.  There is no feedback loop and no data flow from
the library to the application, and thus no \CIVs. Note that this does not
apply to squid, which uses a different API of the libxml2 family.  While this
restrictive pattern cannot suit generic sandboxed API needs, it shows that
there are naturally robust APIs for privilege separation, and remains
applicable to several other scenarios such as (de-)compression or image
processing. We expand more on \CIV-resilient APIs in \S\ref{sec:discussion}.


\insight{
  ``input-only'' APIs are CIV-resilient by design, and can be found is several
  scenarios.
}

\paragraph{Pattern 3 -- Corrupted Data Forwarding}

\begin{listing}
  \begin{minted}{C}
ssize_t send_callback(nghttp2_session *h2,
    uint8_t *mem, size_t length, ...) {
  /* (...), send_underlying = libssl callback */
  written = ((Curl_send*)c->send_underlying)(data,
      FIRSTSOCKET, mem, length, &result);
} /* (...) simplified */
  \end{minted}
  \caption{
Fall-through \CIV in libssl, with curl and libnghttpd2 isolated: curl
transparently forwards corrupted data to libssl.
    \vspace{-0.3cm}
  }
  \label{fig:curlindirectssl}
\end{listing}

In sandbox scenarios, the number of victim components (individual application/libraries/modules that the
untrusted component managed to crash), as shown in \Cref{tab:fuzzingtable}, is
\textgreater 1 for 1/3rd of the cases. This is surprising, since only 4/\numberofscenarios
scenarios have more than one caller component --- 3 of them being FFmpeg.  We
tracked down these observations to one common pattern, where a trusted
component $T_1$ receives corrupted input from an untrusted component $U_1$, and
forwards it to another trusted component $T_2$.  Component $T_2$ thus receives
corrupted data from \emph{trusted} component $T_1$.  Take the example of curl,
which features this \CIV pattern in \Cref{fig:curlindirectssl}. Here, curl's
callback \customtt{send\_callback} ($T_1$) receives corrupted input \customtt{mem}
and \customtt{length} from libnghttpd2 ($U_1$), and transmits it to libssl
($T_2$) via \customtt{send\_underlying}, resulting in a libssl crash. The
untrusted HTTP parsing library thus manages to attack libssl even though the
two libraries do not directly communicate. Intuitively, this pattern motivates
to check as early as possible to avoid unsanitized data spreading.
Unfortunately it is not always possible to perform checks at the untrusted
boundary: $T_1$ is the recipient of the data and may not have
knowledge of the semantics of the corrupted object.  Thus, checks must be
thought with the full system in mind: \emph{it is not because a library only
interfaces with safe or trusted components that it won't receive untrusted
inputs}. This poses the question of ``when to check, and when not to'', to
ensure that no check is missed, while limiting the amount of unnecessary vetting.

\insight{
  the attack surface of a component exceeds interactions with directly reachable
  untrusted components; targeted attacks to non-adjacent components are realistic and checks
  must be thought with the full system in mind.
}

\paragraph{Pattern 4 -- Error-Path \CIVs}

\begin{listing}
	\begin{minted}[escapeinside=||]{C}
static char * ngx_http_block(ngx_conf_t *cf, ...) {
  /* (...) simplified */
  if (module->postconfiguration(cf) != NGX_OK)
    /* cf used in caller after return: */
    return NGX_CONF_ERROR; |\BC{1}|

  if (ngx_http_variables_init_vars(cf) != NGX_OK) |\BC{2}|
    return NGX_CONF_ERROR;
} /* (...) many operations involving cf here */
  \end{minted}
  \caption{
Multi-alteration vulnerability with Nginx modules.
  }
  \label{fig:nginxmultistep}
\end{listing}

About 42\% of the dataset's crashes require more than one alteration to
manifest. In these cases, part of the alterations aim at diverting the control
flow to reach a vulnerable location. We manually studied these multistep
vulnerabilities and found that, in sandbox scenarios, many present a common
pattern where the vulnerability is located in an error-handling path. In this
case, \sys diverts the control flow to the error path (e.g., via a non-zero
return value), so that the application reaches a location that makes use of
another corrupted value.  Take the example of Nginx with the GeoIP module
isolated, as depicted in \Cref{fig:nginxmultistep}.  Nginx calls the module's
post-configuration callback, passing it a pointer to its shared configuration
object \customtt{cf}.  Assume GeoIP corrupts \customtt{cf}.  If the call
succeeds (default behavior in GeoIP), Nginx proceeds with other operations
making use of the configuration object, particularly at \BC{2}, triggering a
crash. Unfortunately, crashes at this stage hide potential for corruption
beyond \BC{2} or in the caller.  Exploiting error-path \CIVs, \sys found that
by corrupting \customtt{cf} and returning an error value from
\customtt{postconfiguration}, further crashes could be uncovered in
\customtt{ngx\_http\_block}'s caller, after the return statement
\BC{1}, avoiding \BC{2}.  Beyond showing that \sys can easily uncover multistep
crashes, this strengthens Pattern 3's observations: checks should be done as
early as possible, ideally before any control flow operation.  \CIV checks
should preempt other functional checks like error code handling to avoid attack
surface multiplication.

\paragraph{Pattern 5 -- Allocator Exposition}

The dataset features 102 allocator corruption bugs affecting a total of 14
scenarios, with 77 bugs coming from GPG with libgcrypt. We investigated this
\CIV class and reduced it to two related patterns, where (1) corrupted
data flows reach parameters of allocators calls, or (2) trusted components expose
untrusted components with a direct window to their allocator. In the first case
we observe numerous DC1-corrupted pointers reaching \customtt{free()},
DC2-corrupted integers reaching \customtt{malloc()}, as well as cases within the
libc due to DC3-corrupted \customtt{FILE*} pointers. The second case appears in
GPG with libgcrypt, and almost systematically in sandbox scenarios with module APIs (HTTPd,
Nginx).  Here, the application presents the sandboxed component with a
malloc-like API that allows it to allocate objects with the
application's memory allocator. This results in numerous DC1-2 \CIVs and
explains the peak in \customtt{void*} related bugs for GPG in
\Cref{fig:typesvuln}.  Allocator exposition is achieved with the intention to
achieve higher performance (custom memory allocator), for introspection
reasons (statistics, or record log information), or for
correctness reasons (allocation/freeing is spread over both components).
Vulnerabilities arising from this pattern are high impact: they allow
malicious compartments to trigger arbitrary use-after-frees, heap Feng
Shui~\cite{Wang2021}, and other allocator exploitation techniques. Since this
pattern also defeats compartment heap separation, it is likely that DL2 \CIVs
will arise too. In all cases, allocator exposition is very difficult to get
right in compartmentalized contexts, and should be avoided where possible.
Removing allocator exposition can be straightforward if performed for
performance reasons only, but may require significant redesign of the API if
done for introspection or correctness reasons.

\insight{
  cross-compartment memory management is hard \& leads
  to exploitable CIVs. Fixes may imply API redesign.
}

\subsection{Security Impact of \CIVs}

At the highest level, our impact analysis confirms that \CIVs are particularly
critical from a security standpoint.  We find that more than 75\% of scenarios
studied present at least one write vulnerability, and all safebox scenarios
present at least one arbitrary read, which defeats attempts to protect secrets.
We present an overview of impact results from \Cref{tab:fuzzingtable}, before
focusing in depth on a selection of bugs from the data set.

As visible in \Cref{tab:fuzzingtable}, the distribution of impact is clearly in
favor of read vulnerabilities, \customtt{NULL} dereferences, followed by write,
allocator corruption, and code execution. This follows the distribution of
typical patterns in applications. Most commonly, a victim component reads data
referenced by a pointer provided by a malicious component (resulting in read
vulnerabilities), or use a corrupted integer within a check (e.g.
a success code) before accessing internal data (resulting in \customtt{NULL}
dereferences if the application reads unallocated/uninitialized data with the
call returning a success code). Less commonly, a victim write to a
pointer provided by a malicious component (write impact). Memory allocator corruption bugs most
commonly happen via pattern 5 of \S\ref{subsec:patterns}, or when size
parameters flow from an interface to an allocation site. The least common
impact is execute, typically resulting from the victim executing a
callback passed by a malicious component.

We find that more than \percentageofarbitraryreadvulnerabilities\% of read and
\percentageofarbitrarywritevulnerabilities\% of write vulnerabilities are
arbitrary, as well as half of execute \CIVs.  Thus, in the absence of
countermeasures, if a subverted component can perform illegal R/W/X operations
outside its compartment through APIs, it is likely to be able to do so at
any address. Further, even though the proportion of execute impact is low (8/\numberofscenarios
scenarios), it is probable that attackers will be able to mount attacks with
arbitrary R/W \CIVs to reach code execution. Next, we illustrate
these observations with an analysis of concrete bugs from the dataset.

\subsubsection{Case Study: OpenSSL Key Extraction}

\begin{listing}
  \begin{minted}{C}
// CIV 1: option setting API leads to arbitrary R/W
ulong SSL_CTX_set_options(SSL_CTX *ctx, ulong op) {
  return ctx->options |= op;
}

// CIV 2: cross-API object SSL_CTX with function
// pointers leads to arbitrary execution
SSL *SSL_new(SSL_CTX *ctx) {
  /* ... */
  s->method = ctx->method;
  /* ... */
  if (!s->method->ssl_new(s)) // arbitrary execution
    goto err;
} /* ... */
  \end{minted}
  \caption{
Two libssl \CIVs leading to arbitrary read, write, and execute impact. Both
functions are exposed to the application.
  }
  \label{fig:libsslcivs}
\end{listing}

\begin{listing}
  \begin{minted}{C}
void aesni_ecb_encrypt(const uchar *in, uchar *out,
  size_t length, const AES_KEY *key, int enc);
  \end{minted}
  \caption{
Prototype of internal encryption API from (I2), vulnerable to key extraction \CIVs.
  }
  \label{fig:libcryptocivs}
\end{listing}

OpenSSL is a popular compartmentalization target, being both high-risk
(high-complexity, shipped in network-facing applications) and sensitive
(holding secrets).  We count at least 8 studies safeboxing it~\cite{Gu2022,
Sartakov2021, Gudka2015, Bauer2021, Agadakos2022, Im2021, Litton2016,
VahldiekOberwagner2019} with attempts going beyond that of academic
research~\cite{Reisinger2014}.  Safeboxing approaches typically either (I1)
compartmentalize OpenSSL in full and isolate at the libssl API, or (I2)
compartmentalize at the libcrypto internal API of key-interacting primitives.
I2 is viewed as more robust because of the reduced TCB and the ability to
tackle intra-libssl bugs such as Heartbleed~\cite{Litton2016}.

We applied \sys to both interfaces. \emph{In all cases, we are able to extract
keys out of the safebox leveraging a single \CIV uncovered by our fuzzer}; we
present three of them.

\Cref{fig:libsslcivs} illustrates two \CIVs found for I1, where libssl is
safeboxed as a whole.  The first \CIV affects libssl's primitives to set/get
SSL options, part of the official libssl API.  Here, an untrusted caller
compartment can control \customtt{ctx} and \customtt{op}, enabling for
arbitrary read/write via the bitmask.  In the second \CIV, libssl executes
callbacks provided by an untrusted object of type \customtt{SSL\_CTX}, one of
the three key vulnerable structures highlighted by \Cref{fig:typesvuln}. These
callbacks are very common across the libssl API and result in arbitrary code
execution. Both vulnerabilities can easily be leveraged to extract the key.

\Cref{fig:libcryptocivs} illustrates a \CIV found for I2, where keys are
isolated at the internal interface. In this encryption primitive, callers
control the \customtt{in} and \customtt{out} pointers, along with the key location
\customtt{key}. By pointing \customtt{in} to the key and pointing \customtt{key} to a
known value, attackers can either cryptanalyze the key out of \customtt{out}, or
use the decryption function to extract the key.

There are systematic problems that make robust safeboxing at the libssl API
difficult. This large API makes heavy use of state structs such as
\customtt{SSL*} or \customtt{SSL\_CTX*}. Sanitizing such structures is hard; it
is likely that, even provided counter-measures from \S\ref{sec:taxonomy}, the
result will approach that of a rewrite of OpenSSL.  Safeboxing at
I2 is less complex but still requires redesign: encryption and
decryption primitives must be made stateful to store the location of valid
keys, checking that input and output buffers do not overlap key locations. Key
creation and loading must also be carefully validated.

\insight{
  \CIVs make it simple to extract SSL keys from an unmodified API safebox.
  Robust SSL key safeboxing requires redesign of the key API
  into a stateful entity.
}

\subsubsection{Case Study: Sudo, Impact Beyond \CIVs}

\begin{listing}
  \begin{minted}{C}
int sudo_passwd_verify(struct passwd *pw, char *pass,
  sudo_auth *auth, struct sudo_conv_callback *cb) {
  /* ... abbreviated ... */
  sav = pass[8]; // read CIV
  pass[8] = '\0'; // write CIV
} /* ... abbreviated ... */
  \end{minted}
  \caption{
    sudo CIV and CVE-2022-43995 manifesting when passed
    password with length below 8 Bytes.
  }
  \label{fig:sudociv}
\end{listing}

Sudo is a strong target for compartmentalization, being high-risk ($>$100K~LoC,
many features) and sensitive (exploits lead to system privilege
escalation). We considered several scenarios, one of them safeboxing the
authentication API, which manages password verification. Here, \sys found 5
\CIVs, with read and \customtt{NULL} dereference impact.  Investigating them,
we realized that one \CIV, shown in \Cref{fig:sudociv}, actually features R/W
impact, and is reachable from user external input, i.e., it is also a
vulnerability in non-compartmentalized contexts.  This decade old issue
manifests when users enter small passwords and was assigned CVE-2022-43995
after we reported it.

\subsubsection{Case Study: Nginx Master/Worker Interface \CIV}

We studied the applicability of \sys to other compartmentalization models such
as the Nginx master/worker manual separation. Here we assume that a worker has
been compromised (e.g., from the network), and attempts to escalate to master
privilege level.  In this model we found a decade-old \CIV that allows a worker
to trigger memory corruption in the master\footnote{\nginxcivurl}. The vulnerability affects a
reliability feature of Nginx: when a worker crashes, the master forcibly
unlocks shared memory mutexes hold by the worker to prevent deadlock. A
malicious worker may corrupt the mutex before crashing itself to force the
master to dereference a crafted pointer. This particular \CIV is low impact due
to control constraints in the mutex unlocking routine -- bytes will only be
overwritten if they match the worker's PID. Nevertheless, \CIVs at such
interfaces present a real risk: less constrained bugs are realistic and may
pose a real privilege escalation threat.

\insight{
  \CIVs also affect production-grade software and may be leveraged
  to mount privilege escalation attacks.
}

\subsection{Conclusions}

We stressed that \CIVs widely affect unmodified software, but in varying
proportions (Q1).  Factors are structural; we elaborated on them with 5 central
patterns and insights (Q2). We illustrated that API redesign will be necessary
in many cases to achieve robust least-privilege enforcement (Q3).  Finally, we
showed that \CIVs are impactful, exploitable, and elaborated with case studies
on popular compartmentalization targets (Q4).  Drawing from this, we discuss
how to design interfaces that are by conception more \CIV-resilient in
\S\ref{subsec:design-for-comp}.

\section{(Re-) Designing Interfaces for Distrust}
\label{sec:discussion}
\label{subsec:design-for-comp}

We showed that interfaces are not equally affected by \CIVs because of
interface design patterns. Next, based on previous sections, we discuss
interface patterns that \emph{reduce} compartmentalization complexity, and how
to leverage them to design strong compartment boundaries, or refactor existing
ones. These patterns do not eliminate the need for \CIV
countermeasures as detailed in \S\ref{sec:taxonomy}; in their absence, these
patterns reduce the number of \CIVs, and in the presence of countermeasures,
these patterns help palliate their limitations. When refactoring, many of the
items listed below require major software redesign. We believe it is a necessary
price to pay to obtain firm safety guarantees from compartmentalization.

\subsubsection{Resources (memory, handles) must be clearly segregated}

Memory ownership must be clearly defined, with each component responsible for
allocating and freeing memory in their region: components must not rely on an
another component's memory allocator (see Pattern 5, \S\ref{subsec:patterns}).
Similarly, system resource handles (or handles to any third-party-managed
resources) must not be shared.  Take the example of \customtt{FILE*}: when
shared, it is hard to determine who should release the handle and when,
requiring complex, ad-hoc, and error-prone virtualization~\cite{Bauer2021,
Im2021, Narayan2020}. Instead, components should acquire and release their own
handle: e.g., for \customtt{FILE*}, components should exchange file paths and
call \customtt{open()} on their own.

\subsubsection{Copy API-crossing objects}

Shared objects must be systematically copied to avoid TV3 \CIVs: it is very
hard to safely use objects that can be concurrently modified by malicious
compartments.  More generally, concurrent usage of objects across compartment
boundaries should be avoided as well, as it introduces the need for
cross-compartment synchronization, which in turn opens for TV2 \CIVs.

\subsubsection{Simplify API-crossing objects}

Compartment interfaces must not expose data that cannot be safely checked.
This includes state information, which is hard to protect in the general case
(Pattern 1, \S\ref{subsec:patterns}). In such cases, a layer of indirection can
be added so that the object is not accessed directly, but through a set of
primitives that can assert the safety of individual operations.  If this is not
possible, the interface is probably not a good compartmentalization boundary in
the first place.

\subsubsection{Trusted-components allocates}

When a trusted component is passed a pointer to a buffer allocated by another
component, it needs to either trust that component, or verify the pointer
(which only privileged monitors can perform as it requires knowledge of the memory
layout). Take the example of C-style strings: if a sandboxed callee
allocates and returns through an API a string pointer, a trusted caller needs
to verify the pointer's validity and the \customtt{NULL}-termination of the
string. This problem can be eliminated by applying a \emph{trusted-component
allocates} policy, i.e., \emph{caller-allocates} in sandbox scenarios, and \emph{callee-allocates} in
safebox scenarios. If the trusted component allocated
the string buffer, it knows the maximum size of the string and can safely check
for \customtt{NULL}-termination.  In the case of mutual
distrust, the involvement of a privileged monitor is necessary.

\subsubsection{Trusted interface functions must be thread-safe}

When a trusted compartment $C_t$ exposes a function $f_t$ (API function for
safeboxes, callbacks for sandboxes) to an untrusted compartment $C_u$,
$C_t$ enables $C_u$ to interfere with its control flow at any time. For
example, $C_u$ may interleave multiple calls to $f_t$ and other API functions
to trigger TV1 \CIVs in $C_t$. Even when calls to $C_t$ functions are
serialized, these may perform callbacks back to $C_u$ that will allow it to
interleave other calls to $C_t$ (a behavior that we observed with ImageMagick
and libpng).  Thus, trusted compartment functions must be designed
\emph{thread-safe} to support any concurrent calling. Alternatively, trusted
interface calls should be strictly serialized and run to completion (no
callbacks), a rather restrictive model.

\subsubsection{Trusted interface functions must define \& enforce ordering
requirements}

Similarly, if trusted interface elements $f_1\cdots{}f_n$ have ordering
requirements, then these must be clearly stated and enforced to further tackle
TV1 \CIVs. This may require safeboxed libraries to become stateful, where they
previously relied on invoking undefined behavior if the caller did not respect
ordering. When asynchronous behavior is suitable, event-loop-based designs may
allow interface designers to shift as much control-flow leverage as possible
out of the hand of attacker by processing the core of the callback in the main
loop, in a way that is consistent with other external inputs (similarly to
signal processing).

\subsubsection{No sharing of uninitialized data}

API-crossing uninitialized data must be systematically zero-ed to avoid DL1-2
\CIVs (\S\ref{subsec:info-exposure}).  Even sharing of properly checked objects
can be unsafe if they have not been zero-ed at initialization, since
compiler-added padding might remain uninitialized. Where applicable, zeroing
should be compiler-enforced.

\subsubsection{\CIV checks first}

As soon as one allows untrusted data to propagate unchecked through a
compartment, it becomes hard to ensure that all checks are properly performed
down the line (Pattern 4, \S\ref{subsec:patterns}), and encourages duplication
of non-trivial checks, maximizing the likelihood of errors in present and
future versions of the software. Worse, untrusted data might not even be used
within but simply flow through a compartment to be used in another one, which
might have variable trust assumptions on the compartment feeding it data
(Pattern 3, \S\ref{subsec:patterns}). Copy and checks should therefore be
performed on the data as soon as possible after crossing the API, and preempt
all other functional checks.

\section{Related Works}\label{sec:related-works}

\paragraph{Finding API Vulnerabilities}

DUI detector ~\cite{Hu2015} leverages static binary analysis, symbolic
execution and dynamic taint analysis to detect pointer dereferences made by a
security domain under the influence of another through an interface. Due to
performance and scalability issues (emulation and symbolic execution), such
approaches are hard to scale to large programs, large interfaces, large numbers
of programs, or, as is the case here, high bug counts. Further, compiler-based
static approaches are unsuited to scenarios like OpenSSL, where safeboxed
components are implemented in pure assembly files. Other studies such as Van
Bulck et al.~\cite{VanBulck2019}, focusing on Trusted Execution Environment
(TEE) runtimes, take a manual approach to identify interface vulnerabilities.
Being manual, such approaches are limited in scope. In-memory fuzzing allows us
to be faster and more scalable than static and manual approaches, enabling for
a larger-scale \CIV study. Fuzzing yields a subset of all \CIVs
present at an interface, which is a suitable limitation in our case.

Classical system call fuzzing~\cite{SYZKALLER} searches for kernel
vulnerabilities at the system call API. This corresponds to \emph{one specific
safebox scenario} where the compartment API is the system call API. \sys is much
more general, targeting arbitrary sandbox/safebox scenarios at arbitrary APIs.
Similarly to system call fuzzing, Emilia~\cite{Cui2022} fuzzes for
Iago~\cite{Checkoway2013} vulnerabilities, hooking at system calls and altering
their return values to simulate a malicious kernel, corresponding to a sandbox
model. Here too, \sys is much more general as it 1) can hook into arbitrary
interfaces; 2) supports bidirectional fuzzing (sandbox/safebox); and 3)
fuzzes the \emph{full compartment attack surface} (callbacks, return
values, shared data, function arguments) -- whereas Emilia only fuzzes return
values.

\paragraph{Finding API Misuses}

APISan~\cite{Yun2016} studies existing software to infer semantic usage
information for a given API (e.g., semantic relation on arguments/functions).
Using that information, it searches for deviations to detect possible API
misuses at the source code level.  Such an approach is not suited to detect
\CIVs. First, \CIVs can be present even when the semantics of an API are
respected in the code: at runtime, a malicious compartment with code execution
abilities can manipulate the program's execution in a way that does not respect
the API semantics. Second, even if API semantics could be fully enforced at
runtime, most \CIVs would remain undetected because the semantics of unmodified
APIs are generally unsuited to distrust, as we show in the paper.
Nevertheless, as we highlight in \S\ref{sec:taxonomy}, APISan's ability to
infer API semantics may be leveraged to determine enforcement policies,
provided a large enough set of API usage samples (usually available for
popular APIs).

\paragraph{In-Memory Fuzzing}

Unlike conventional fuzzing approaches that inject malformed data through a
program's input channels (e.g., network), in-memory fuzzing~\cite{Sutton2007}
moves the fuzzer within the target using process instrumentation techniques.
\sys is an in-memory fuzzer specialized for \CIV fuzzing. \sys mainly differs
from existing in-memory fuzzers~\cite{Sutton2007, Serebryany2016, Manes2021} in
that it 1) fuzzes in both ways (sandbox and safebox) -- whereas existing
in-memory fuzzers mostly correspond to safebox fuzzing, and 2) targets a
different attack surface, the \textit{compartment} attack surface -- which,
unlike usual in-memory fuzzers, also includes callbacks, return values, etc. To
our knowledge, we are the first to use in-memory fuzzing with the goal of
studying \CIVs in unmodified software.

\paragraph{Interface-Aware Compartmentalization Frameworks}

Compartmentalization frameworks provide a variable degree of support for
protecting security domain interfaces. The vast majority of modern
compartmentalization frameworks~\cite{Wu2012, VahldiekOberwagner2019,
Hedayati2019, Schrammel2020, Liu2017, Koning2017, Park2019, Bauer2021,
Sartakov2021, Lefeuvre2021, Lefeuvre2022, Agadakos2022} do not achieve more
than basic ABI-level interface sanitization at security domain crossing, such
as switching the stack and clearing registers.  Combined with the fact that
most also rely on relatively coarse-grain shared memory-based communication for
performance reasons, this opens up a wide range of \CIVs and was one of our
motivations to develop \sys.

RLBox is a sandboxing framework for untrusted C++ software components. RLBox
sanitizes sandbox data flow in a partially automated way: using static analysis
and C++ type information, the framework can add certain checks automatically.
When not possible, RLBox outputs compiler errors to require human intervention.
Similarly, SOAAP~\cite{Gudka2015} relies on code annotations and employs static
analysis to flag possible data leaks.  Both approaches are prone to human error
due to manual effort.  The CHERI~\cite{Watson2015} hardware memory capability
model promises strong and efficient compartmentalization by extending RISC ISAs
with capability instructions. Certain CHERI features (e.g., unforgeable
pointers/capabilities, byte-level memory sharing) eliminate or mitigate some
classes of \CIVs.  Nevertheless, CHERI is still a
prototype~\cite{ARMMorello2020}.  We discuss the benefits and limitations of
all three systems in \S\ref{sec:taxonomy}.

Several TEE runtimes have been proposed~\cite{OE_SDK, Tsai2017, Priebe2019,
ASYLO, FORTANIX_EDP} to transparently shield enclaves from the outside world by
maintaining a secure interface. However, as demonstrated by several
studies~\cite{VanBulck2019, Cui2022} this cannot eliminate all \CIVs,
motivating fuzzers such as Emilia~\cite{Cui2022} and \sys. Before TEEs,
sandboxing frameworks protecting applications from a malicious OS such as
InkTag~\cite{Hofmann2013} and MiniBox~\cite{Li2014} attempted to prevent Iago
attacks by vetting/managing the memory mappings requests made by the protected
program.

\section{Conclusion}\label{sec:conclusion}

Breaking down monolithic software into compartments without reasoning about
newly created interfaces leads to \CIVsfull. This paper presented an in-depth
study of \CIVs.  We proposed \sys, an in-memory fuzzing approach to investigate
\CIVs and their impact in compartmentalized software. Applying it to
\numberofapps applications and \numberoflibs libraries, we uncovered a large
data-set of \numberofcrashes \CIVs from which we extracted numerous insights on
the prevalence of CIVs, their causes, impact, and the complexity to address
them: we confirmed how important CIVs should be to compartmentalization
research, and highlighted how API design patterns influence their prevalence
and severity. We concluded by stressing that addressing these problems is more
complex than simply writing a few checks, proposed guidance on
compartmentalization-aware interface design and adaptation, and motivated for
more research towards systematic CIV detection and mitigation.  We open-sourced
code and data: \urlsys.

\section*{Acknowledgements}

We thank the anonymous reviewers for their insights. We are also grateful to
David Chisnall and Istvan Haller for their insightful feedback. This work was
partly funded by a studentship from NEC Labs Europe, a Microsoft Research PhD
Fellowship, the UK’s EPSRC grants EP/V012134/1 (UniFaaS), EP/V000225/1
(SCorCH), the EPSRC/Innovate UK grant EP/X015610/1 (FlexCap), the EU H2020
grant agreements 871793 (ACCORDION) and 758815 (CORNET), and the NSF CNS
\#2008867, \#2146537, and ONR N00014-22-1-2057 grants.

\bibliographystyle{plain}
\bibliography{bib}

\end{document}